\documentclass[aps,pra,twocolumn,superscriptaddress]{revtex4}

\usepackage{amsmath}
\usepackage{amssymb}
\usepackage{mathrsfs}
\usepackage{verbatim}
\usepackage{epsfig}
\usepackage{color}
\usepackage{soul}
\usepackage{ulem}
\usepackage[bookmarks]{hyperref}
\usepackage{graphicx}

\graphicspath{{.}{./EPS/}}

\newcommand* {\bra}[1]{\ensuremath{\langle {#1} |}}
\newcommand* {\ket}[1]{\ensuremath{| {#1} \rangle}}

\begin{document}

\title{Quantitative characterization of several entanglement detection criteria}

\author{A. Sauer}
\affiliation{Institut f\"{u}r Angewandte Physik, Technische Universit\"{a}t
Darmstadt, D-64289 Darmstadt, Germany}
\email{alexander.sauer@physik.tu-darmstadt.de}

\author{J. Z. Bern\'ad}
\affiliation{Peter Gr\"unberg Institute (PGI-8), Forschungszentrum J\"ulich, D-52425
J\"ulich, Germany}
\email{j.bernad@fz-juelich.de}

\date{\today}

\begin{abstract}
Quantitative characterization of different entanglement detection criteria for bipartite systems is presented. We review the implication sequence of these criteria and then numerically estimate volume ratios between 
criteria non-violating quantum states and all quantum states. The numerical approach is based on the hit-and-run algorithm, which is applied to the convex set of all quantum states embedded into a Euclidean vector space
of  the Hilbert-Schmidt inner product. We demonstrate that reduction, majorization, and the R\'enyi-entropy-based criteria are very ineffective compared to the positive partial transpose. In the case of the 
R\'enyi-entropy-based criterion, we show that the ratio of detectable entanglement increases with the order of the R\'enyi entropy.   
\end{abstract}

\maketitle

\section{Introduction}
\label{I}

In the last couple of decades quantum information science has seen  an explosive development and entanglement has been identified as the main physical resource in various applications 
\cite{Peres97, Nielsen, Bengtsson17, Bruss}. In response, several criteria have been proposed characterizing separable versus entangled states. Entanglement detection criteria, which are based on
entanglement witnesses, positive maps \cite{Horodecki96, Horodecki01} and the projective cross norm \cite{Rudolph, Arveson}, are capable to completely characterize both sets of separable and entangled states, 
but they do not provide fast and simple computational methods. In parallel to these abstract developments, historically other criteria were formulated in the form of simple algebraic tests 
\cite{Terhal, Guhne, Horodecki09, Simnacher}, like the well-known Peres-Horodecki criterion obtained by the partial transposition of density matrices \cite{Horodecki96, Peresp}. These tests are computationally very practical and some of them are even
implemented experimentally \cite{Bovino05, Schmid, Islam, Bartkiewicz17}. However, in most cases they provide only necessary conditions for the separability of the states. Throughout the last two decades,  
relations between these criteria have been found and thus a qualitative ordering is established. Quantitative characterization of the Peres-Horodecki criterion was initiated by Ref. \cite{ZykLew}, however,
regarding the typicality study of those quantum states which violate the other criteria it is still missing. This paper is devoted to the numerical study of this characterization in several bipartite systems. 

In this paper, we assign to every quantum state or density matrix a point in the Euclidean vector space defined by the Hilbert-Schmidt inner product. Thus, the convex set of density matrices is mapped into a convex 
body of the Euclidean vector space. Any entanglement detection criterion separates this convex body into two disjunct subsets, i.e., a density matrix either violates or does not violate the criterion. The volume 
ratio of these subsets is going to give the quantitative characterization of each criterion. To estimate these ratios we employ a numerical approach developed recently by us and based on the hit-and-run (HR) 
algorithm \cite{Sauer}, which realizes a random walk inside the set of density matrices. The HR sampler generates asymptotically and effectively uniformly distributed points over any convex body $K$ and moreover, this is independent 
of the starting point inside $K$ \cite{Smith, Lovasz0, Lovasz}. We start our numerical investigation with the Peres-Horodecki criterion, for which there are now numerous results for the typicality of quantum states with 
positive partial transpose (PPT); see Refs. \cite{Sauer,Slater} and also the references therein. In Ref. \cite{Sauer}, we have studied the typicality of bipartite two-qubit entanglement 
which can be detected by violations of Bell inequalities and the PPT criterion up to $3 \times 3$ (qutrit--qutrit) bipartite quantum systems. Here, we extend the study of the PPT criterion with new estimates for 
$2 \times 5$ (qubit--five-level qudit) and $3 \times 4$ (qutrit--four-level qudit) bipartite
quantum systems. This is followed up by the numerical investigation of the reduction criterion \cite{Cerf, Horodecki2}, the majorization criterion \cite{Majorization}, and criteria based on R\'enyi entropies 
\cite{Renyi}. We show that many of these criteria become less effective in the task of detecting entanglement with the increase of the dimension of bipartite systems. 

The paper is organized as follows. In Sec. \ref{II}, we recall the definitions and the known implications of all criteria being subject to our investigations. 
A brief description of our numerical approach is also presented in Sec. \ref{III}. Numerical results 
for different bipartite systems are discussed in Sec. \ref{IV}.  Finally, Sec. \ref{V} contains our conclusions.

\section{Criteria on separable states}
\label{II}

In this section we give an overview on some of the entanglement detection criteria and their relations to each other. We consider the finite-dimensional Hilbert spaces $\mathbb{C}^n$, where a 
density matrix or quantum state $\rho$ is 
defined as a positive semidefinite matrix acting on $\mathbb{C}^n$ with unit trace
\begin{equation}
 \rho \geqslant 0, \quad \mathrm{Tr} \{\rho\}=1. \nonumber
\end{equation}
A density matrix $\rho_{AB}$ of a bipartite system is defined on the Hilbert space $\mathbb{C}^{n_A} \otimes \mathbb{C}^{n_B}$, where $n_A$ and $n_B$ are the dimensions of 
the subsystems $A$ and $B$. If $\rho^{(A)}$ and $\rho^{(B)}$ are density matrices of $\mathbb{C}^{n_A}$ and $\mathbb{C}^{n_B}$, respectively, then $\rho^{(A)} \otimes \rho^{(B)}$ is called 
a tensor product state. When $\rho_{AB}$ can be written as a convex combination of tensor product states, it is called separable:
\begin{eqnarray}
 \rho_{AB}&=&\sum_k p_k \rho^{(A)}_k \otimes \rho^{(B)}_k, \quad p_k \geqslant 0, \quad \sum_k p_k=1. \label{separable}
\end{eqnarray}
It is clear from these definitions that the set of all density matrices and as well the set of the separable ones are convex. If a density matrix does not have the form \eqref{separable}, it is called an
entangled state. Checking if a quantum state is separable or entangled is a hard problem and therefore several entanglement detection criteria have been proposed. In the following subsections, we review those ones, 
which are subject to our numerical investigations.
\begin{figure}[t!]
   \centering
  \includegraphics[width=0.4\textwidth]{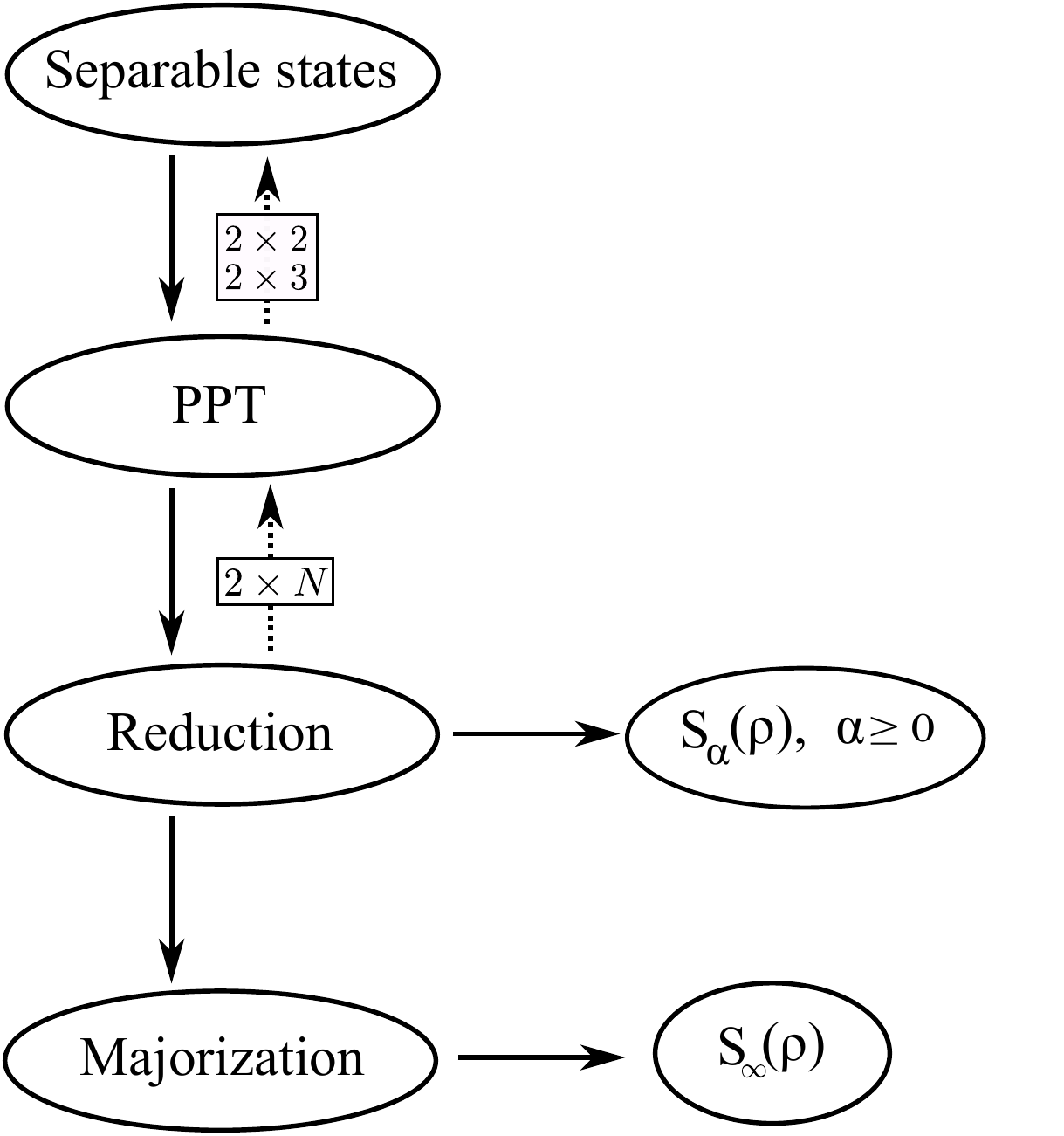}
     \caption{Relations between various entanglement detection criteria.  Arrows denote that a state meeting one criterion will also meet the criterion the arrow is pointing at. Dashed arrows indicate special 
     cases when double implications are possible.}
   \label{fig:criteria}
 \end{figure}

\subsection{Positive partial transpose (PPT)}

A simple, but computationally tractable criterion was found by Peres \cite{Peresp}. Let us consider a finite-dimensional bipartite quantum system with Hilbert space $\mathbb{C}^{n_A}\otimes\mathbb{C}^{n_B}$,
the transposition map $\mathrm{\tau}_{A}$ on $A$, and the identity operation $\mathbb{I}_{B}$ on $B$. Then, the partial transposition map $\rho_{AB} \rightarrow (\mathrm{\tau}_{A}\otimes\mathbb{I}_{B})\rho_{AB}$ 
is defined with respect to the canonical product basis as 
\begin{equation}
\bra{ij} (\mathrm{\tau}_{A}\otimes\mathbb{I}_{B})\rho_{AB} \ket{kl}=\bra{kj} \rho \ket{il}. \nonumber 
\end{equation}
If we apply $\mathrm{\tau}_{A}\otimes\mathbb{I}_{B}$ on a separable density matrix, then we always get a density matrix. This criterion is capable of completely characterizing the set of separable quantum states only 
for $2 \times 2$ (qubit--qubit) and $2 \times 3$ (qubit--qutrit) bipartite systems \cite{Horodecki96} and is independent of the subsystem that is transposed. In larger systems the situation is more involved, 
because there exist entangled states which satisfy the PPT criterion, i.e, the so-called phenomenon of bound entanglement or entangled PPT states \cite{Horodecki98}. 

\subsection{Reduction criterion}

The reduction criterion poses a condition on reductions of the density matrix to the two subsystems. We consider the partial traces to the subsystems $\rho_A = \mathrm{Tr}_B \{\rho_{AB}\} $ 
and $\rho_B = \mathrm{Tr}_A \{\rho_{AB}\}$. Furthermore, we denote the identity matrices on $\mathbb{C}^{n_A}$ and  $\mathbb{C}^{n_B}$ by $I_{n_A}$ and $I_{n_B}$, respectively. All separable and PPT states 
fulfill the condition \cite{Horodecki2}
\begin{equation}
\rho_A \otimes I_{n_B}- \rho_{AB} \geqslant 0  \quad  \text{and} \quad I_{n_A} \otimes   \rho_B- \rho_{AB} \geqslant 0,
\label{eq:red}
\end{equation}
i.e., the left-hand side is always a positive semidefinite matrix. It has been shown that this criterion is identical to PPT for $2\times N$ (qubit--$N$-level-qudit) bipartite quantum systems \cite{Cerf}.

\subsection{Majorization criterion}

In a similar approach to the reduction criterion, one can investigate the eigenvalues of the density matrix and its partial traces.
Let $\vec{\lambda}^\downarrow$ be the vector with coordinates of $\vec{\lambda}$ rearranged in descending order. We say $\vec{\lambda} \in \mathbb{R}^n$ is majorized 
by $\vec{\mu} \in \mathbb{R}^n$ if 
\begin{equation}
 \sum_{i=1}^k \lambda^\downarrow_i \leqslant \sum_{i=1}^k \mu^\downarrow_i, \quad 1 \leqslant k \leqslant n, \nonumber
\end{equation}
and 
\begin{equation}
 \sum_{i=1}^n \lambda^\downarrow_i = \sum_{i=1}^n \mu^\downarrow_i, \nonumber 
\end{equation}
which is denoted by $\vec{\lambda} \prec \vec{\mu}$. If $\vec{\lambda}(\rho)$ is the vector of eigenvalues of $\rho$, then for all separable states \cite{Majorization}
\begin{equation}
\vec{\lambda}(\rho_{AB}) \prec \vec{\lambda}(\rho_{A}) \quad \text{and} \quad \vec{\lambda}(\rho_{AB}) \prec \vec{\lambda}(\rho_{B}),
\label{eq:maj}
\end{equation}
where $\vec{\lambda}(\rho_{A})$ and $\vec{\lambda}(\rho_{B})$ are enlarged by appending extra zeros to equalize their dimensions with the dimension of $\vec{\lambda}(\rho_{AB})$.
In our case all eigenvalues are non-negative and they sum to $1$. Thus, it is sufficient to compute the first $n-1$ pairs of elements.
This criterion is weaker than the reduction criterion, as all quantum states fulfilling the latter also obey the majorization criterion \cite{Hiroshima}.

\subsection{Criterion based on R\'enyi entropies}

For a real number $\alpha>0$ with $\alpha \neq 1$ the R\'enyi entropy of a density matrix $\rho$ is defined as
\begin{equation}
S_\alpha(\rho) = \frac{1}{1-\alpha} \ln \mathrm{Tr} \{\rho^\alpha\} \nonumber.
\end{equation}
For two special cases we have
\begin{equation}
 \lim_{\alpha \to 1} S_\alpha(\rho) = - \mathrm{Tr} \{\rho \ln \rho \}, \nonumber
\end{equation}
i.e., von Neumann entropy, and 
\begin{equation}
 S_\infty (\rho)=- \ln \| \rho \|, \nonumber
\end{equation}
where $\| \cdot \|$  denotes the operator norm. Density matrices are obviously normal matrices and therefore $\|\rho\|$ is the largest eigenvalue of $\rho$. If $\rho$ is separable,
\begin{equation}
S_\alpha \left( \rho_A \right)  \leqslant S_\alpha \left( \rho_{AB} \right) \quad \text{and} \quad  S_\alpha \left( \rho_B \right)  \leqslant S_\alpha \left( \rho_{AB} \right)
\label{eq:entr}
\end{equation}
for $\alpha>0$. It was also shown that the reduction criterion implies all entropy criteria \cite{Vollbrecht}.  If $\rho_{AB}$ fulfills the majorization criterion, then both largest
eigenvalues of $\rho_A$ and $\rho_B$ are greater than or equal to the largest eigenvalue of $\rho_{AB}$, which yields that the entropic criterion with $S_\infty (\rho_{AB})$ is also fulfilled. 
The special case $\alpha=1$ is of particular interest, as only states violating the corresponding condition may be useful for dense coding \cite{Bruss}.
One can consider $\alpha \to 0$ as well, when we have $S_0(\rho)=\ln \operatorname{rank} (\rho)$ where $\operatorname{rank} (\rho)$ is the rank of $\rho$. However, we do not consider this case, because in our numerical approach 
we generate full rank density matrices, so the set of density matrices with at least one zero eigenvalue has measure zero. Finally, in Fig. \ref{fig:criteria} we have sketched an overview of the 
general sequence of implication between the above discussed entanglement detection criteria.   
Other conditions on separable quantum states, e.g., matrix realignment \cite{Chen}, are not discussed here as there is no clear connection to the previously mentioned criteria. 
For example, there are states violating the PPT but not the matrix realignment criterion \cite{Rudolph2} and vice versa \cite{Chen}. However, a combination of both criteria minimizes 
the amount of candidates for separable quantum states \cite{Zhang}. 

\section{Numerical approach}
\label{III}

It is known that the vector space of $n \times n$ matrices with complex entries $M_n(\mathbb{C})$ with the Hilbert-Schmidt inner product is a $n^2$-dimensional Hilbert space, i.e, a Euclidean vector space. In this 
Euclidean vector space, self-adjoint matrices form a subspace, which with the help of normalized generators $T_i$ of the Lie group $SU(n)$ and the unit matrix $I_n/\sqrt{n}$ can be identified with $\mathbb{R}^{n^2}$. It is worth noting 
that there are other possible choices of orthonormal bases, e.g., the Gell-Mann-type basis of $SU(4)$ instead of the basis built up from the Pauli matrices, but they always result  in the same Euclidean structure. 
Other interesting orthonormal bases exist \cite{Bertlmann}, e.g., the Weyl operator basis, though not all are suitable for our approach based on $\mathbb{R}^{n^2}$.

If $A$ is self-adjoint, 
then
\begin{equation}
 A = a_0 \frac{I_n}{\sqrt{n}}+\sum^{n^2-1}_{i=1} a_i T_i. \nonumber
\end{equation}
We are interested in the subset subject to $\mathrm{Tr}\{ A\}=1$, i.e., $a_0=1/\sqrt{n}$. Density matrices lie in this subset and have the form
\begin{equation}
 \rho = \frac{I_n}{n}+\sum^{n^2-1}_{i=1} a_i T_i, \label{eq:densitym} 
\end{equation}
where the $a_i$s have to fulfill $n-1$ conditions based on Newton identities and Descartes' rule of signs \cite{Kimura}. Therefore, these conditions define the boundaries of the convex body of density matrices 
in $\mathbb{R}^{n^2-1}$ and we denote this body by $K$. Furthermore, we consider ${\bf a}=(a_1, a_2, \dots a_{n^2-1})^T \in \mathbb{R}^{n^2-1}$ ($T$ denotes the transposition). In order to estimate the volume ratios 
between states which do not violate an entanglement detection criterion and all states, we consider a random walk in $K$. This is 
done by the HR algorithm:
\begin{itemize}
 \item 1. Initialize with ${\bf a}^{(j)}$ and set the iteration counter $j=1$. We always pick ${\bf 0}$ or the zero vector, i.e., the maximally mixed state.
 \item 2. Generate a random direction ${\bf d}^{(j)}$ according to a uniform distribution on the unit $(n^2-1)$-dimensional hypersphere.
 \item 3. Let $r=2\sqrt{n-1}/\sqrt{n}$ and set $I=[-r,r]$
 \item 4. Generate $\lambda$ uniformly within the interval $I$.
 \item 5. If ${\bf a}^{(j)}+\lambda {\bf d}^{(j)} \in K$, then ${\bf a}^{(j+1)}={\bf a}^{(j)}+\lambda {\bf d}^{(j)}$ and go back to Step 2. Otherwise, set the interval to $[\lambda, r]$ or $[-r,\lambda]$ such 
 that zero is included and return to step 4.
\end{itemize}
The Markov chain underlying this algorithm converges in the sense of total variation distance to the uniform stationary distribution in polynomial time \cite{Lovasz0}, called also the mixing time. 
If we start from a point at distance $l$ from the boundary of $K$, then based on the results of Lov\'asz and Vempala in Ref. \cite{Lovasz} HR mixes 
in $O(n^4 \ln^3 (n/l))$ steps. When the convex body is in the so-called near-isotropic position \cite{Kannan}, then at least $O(n^3)$ steps are required \cite{Lovasz}. The center of mass of $K$ is the origin and 
therefore we expect that a sample size between $O(n^3)$ and $O(n^4 \ln^3 n)$ is enough to ensure an almost uniform distribution of density matrices. Estimates and their standard deviations are obtained in the same way
as we did in Ref.\cite{Sauer}.

\section{Results}
\label{IV}

In this section we investigate numerically  all the criteria presented in Sec. \ref{II}. The aim is to provide estimates and their standard deviations for volume ratios $R$ between quantum states not violating a 
criterion and the whole set of quantum states. All bipartite quantum systems which are denoted as $n_A \times n_B$ refer to the most general form of quantum states composed of an $n_A$-dimensional qudit and an 
$n_B$-dimensional qudit. A density matrix of such a quantum state can then be written in the form
\begin{eqnarray}
&&\rho_{AB} = \frac{I_{n_A \times n_B}}{n_A n_B} +\frac{1}{\sqrt{n_B}} 
 \sum^{n^2_A-1}_{i=1} \tau^{(A)}_i \, T^{(A)}_i  \otimes I_{n_B} \label{full} \\
&&+\frac{1}{\sqrt{n_A}} \sum^{n^2_B-1}_{j=1} \tau^{(B)}_j \,I_{n_A} \otimes T^{(B)}_j + \sum_{i,j} \nu_{i,j} \,T^{(A)}_i  \otimes  T^{(B)}_j, \nonumber
\end{eqnarray}
where $T^{(A)}_i$ and $T^{(B)}_j$ are the normalized generators of the Lie group $SU(n_A)$ and $SU(n_B)$, respectively.  In addition we look at some interesting subsystems for qubit--qubit and qubit--qutrit systems.
In the case of a qubit--qubit or $2 \times 2$ system we have the well-known Bell-diagonal states with 
\begin{equation}
    \rho_{\rm BD} = \frac{I_4}{4}+\frac{1}{2}\sum_{i=x,y,z}\,a_i\,\sigma^{(A)}_i\otimes
    \sigma^{(B)}_i, 
    \label{eq:belldiag}
\end{equation}
where $\sigma_i$ are the Pauli matrices. Then, we study the so-called X-states \cite{Rau}:
\begin{equation}
\rho_{\rm X} = \frac{I_4}{4} + \frac{1}{2}\sum_{i=1}^7 a_i\,T_i
\label{eq:Xs}
\end{equation}
with 
\begin{eqnarray}
T_1&=&\sigma^{(A)}_z\otimes I^{(B)}_2, \, T_2=I^{(A)}_2 \otimes \sigma^{(B)}_z, \, T_3=\sigma^{(A)}_x\otimes \sigma^{(B)}_x, \nonumber \\
T_4&=&\sigma^{(A)}_x\otimes \sigma^{(B)}_y, \, T_5=\sigma^{(A)}_y\otimes \sigma^{(B)}_x, \, T_6=\sigma^{(A)}_y \otimes \sigma^{(B)}_y, \nonumber \\
T_7&=&\sigma^{(A)}_z\otimes \sigma^{(B)}_z. \nonumber
\end{eqnarray}
Another interesting family of states is the so-called rebit--rebit states, i.e., real valued two-qubit states:
\begin{equation}
\rho_{\rm RR} = \frac{I_4}{4} + \frac{1}{2}\sum_{i=1}^9 a_i\,T_i.
\label{eq:rebit}
\end{equation}
Here, the $T_i$s are chosen in such a way that the entries of $\rho_{\rm RR}$ are real:  
\begin{eqnarray}
T_1&=&I^{(A)}_2 \otimes \sigma^{(B)}_x, \, T_2=I^{(A)}_2 \otimes \sigma^{(B)}_z, \, T_3=\sigma^{(A)}_x\otimes I^{(B)}_2, \nonumber \\
T_4&=&\sigma^{(A)}_z\otimes I^{(B)}_2, \, T_5=\sigma^{(A)}_x\otimes \sigma^{(B)}_x, \, T_6=\sigma^{(A)}_x \otimes \sigma^{(B)}_z, \nonumber \\
T_7&=&\sigma^{(A)}_y\otimes \sigma^{(B)}_y, \, T_8=\sigma^{(A)}_z \otimes \sigma^{(B)}_x, \, T_9=\sigma^{(A)}_z \otimes \sigma^{(B)}_z. \nonumber
\end{eqnarray}
In the case of a qubit--qutrit or $2 \times 3$ system we also investigate the ratios for the following subsystems \cite{Sauer}:
\begin{eqnarray}
 (i) \quad \rho_{I} &=&\frac{{I_6}}{6}+\frac{1}{2}\sum_{i=1}^4 \nu_{x,i}\,\sigma^{(A)}_x \otimes \gamma^{(B)}_i \label{eq:qbqt-d12} \\
       &+& \frac{1}{2}\sum_{i=1}^4 \nu_{y,i}\,\sigma^{(A)}_y \otimes \gamma^{(B)}_i + \frac{1}{2}\sum_{i=1}^{4} \nu_{z,i}\,\sigma^{(A)}_z \otimes \gamma^{(B)}_i, \nonumber \\
 (ii) \quad \rho_{II} &=&\frac{{I_6}}{6}+\frac{1}{2}\sum_{i=1}^8 \nu_{x,i}\,\sigma^{(A)}_x \otimes \gamma^{(B)}_i  \label{eq:qbqt-d24} \\
    &+& \frac{1}{2}\sum_{i=1}^8 \nu_{y,i}\,\sigma^{(A)}_y \otimes \gamma^{(B)}_i + \frac{1}{2}\sum_{i=1}^{8} \nu_{z,i}\,\sigma^{(A)}_z \otimes \gamma^{(B)}_i, \nonumber 
\end{eqnarray}
where $\gamma_i$ are the Gell-Mann matrices. Furthermore, we denote by $d$ the dimension of the Euclidean vector space, e.g., in the case of Bell-diagonal states $d=3$, but for two qutrits $d=3^2 \cdot 3^2-1=80$. 
The calculations of volumes of different sets of quantum states are based on the Lebesgue measure in $\mathbb{R}^d$.

\begin{table*}[t!]
 \centering
\begin{tabular}{|c | c | c  c c   c c | c | } 
\hline
 &    &  PPT  &  Reduction  & Majorization &  $S_1(\rho)$  &  $ S_\infty(\rho) $ & Sample size\\ 
\hline
$2 \times 2$ & Bell-diagonal & \uwave{0.49997(10)} & \uwave{0.49997(10)} & \uwave{0.49997(10)} & 0.958559(52) & \uwave{0.49997(10)} &$1 \times 10^8$\\ 
$2 \times 2$ & X-state & \underline{0.39990(14)} & \underline{0.39990(14)} & \uwave{0.64690(16)} &  0.977187(55) & \uwave{0.64690(16)}& $1 \times 10^8$\\ 
$2 \times 2$ & rebit--rebit & \underline{0.45317(17)} & \underline{0.45317(17)} & \uwave{0.80822(17)} &  0.992395(38) & \uwave{0.80822(17)}& $1 \times 10^8$\\ 
$2 \times 2$ & general & \underline{0.24244(17)} & \underline{0.24244(17)} & \uwave{0.78464(24)} & 0.995278(36) & \uwave{0.78464(24)}& $1 \times 10^8$\\ 
$2 \times 3$  & (i) & \uwave{0.19384(29)} & \uwave{0.19384(29)} & \uwave{0.19384(29)} &  0.9999625(55) & \uwave{0.19384(29)}& $2.1 \times 10^7$\\ 
$2 \times 3$  & (ii) & \uwave{0.02226(16)} & \uwave{0.02226(16)} & \uwave{0.02226(16)} &  0.999933(12) & \uwave{0.02226(16)}& $1.1 \times 10^7$\\ 
$2 \times 3$  & general & \underline{0.02673(13)} & \underline{0.02673(13)} & 0.86168(67) &  0.999909(18) & 0.86746(66)& $3.1 \times 10^7$\\ 
$2 \times 4$ & general  & \underline{0.001229(60)} & \underline{0.001229(60)} & 0.8824(23) &1  & 0.8877(22) & $2.5 \times 10^7$\\ 
$3 \times 3$ & general  & 0.0001058(85) & 0.6470(21) & \uwave{0.99528(35)} &  1 & \uwave{0.99528(35)}& $1.2 \times 10^9$\\  
$2 \times 5$ & general  & \underline{0.00002606(88)} & \underline{0.00002606(88)} & 0.89974(38) &  1 & 0.90416(36)& $4 \times 10^8$\\ 
$3 \times 4$ & general  & $(6.0 \pm 4.0) \cdot 10^{-8}$ & 0.5743(11) & \uwave{0.99861(58)} &  1 & \uwave{0.99861(58)}& $3.5 \times 10^8$\\
\hline
\end{tabular}
\caption{Volume ratios between quantum states fulfilling different criteria and all corresponding quantum states. In the case of $S_1(\rho)$, i.e., the von Neumann entropy-based criterion, and systems larger than $2\times 3$, 
the results are inconclusive, because no states violating the criterion were sampled. The statistical errors are given in brackets after the obtained value, e.g., $0.49997(10) \equiv 0.49997 \pm 0.00010 $. Identical numerical estimates in the same row are highlighted.}
\label{tab:ratios}
\end{table*}

\begin{figure}[h!]
   \centering
  \includegraphics[width=0.5\textwidth]{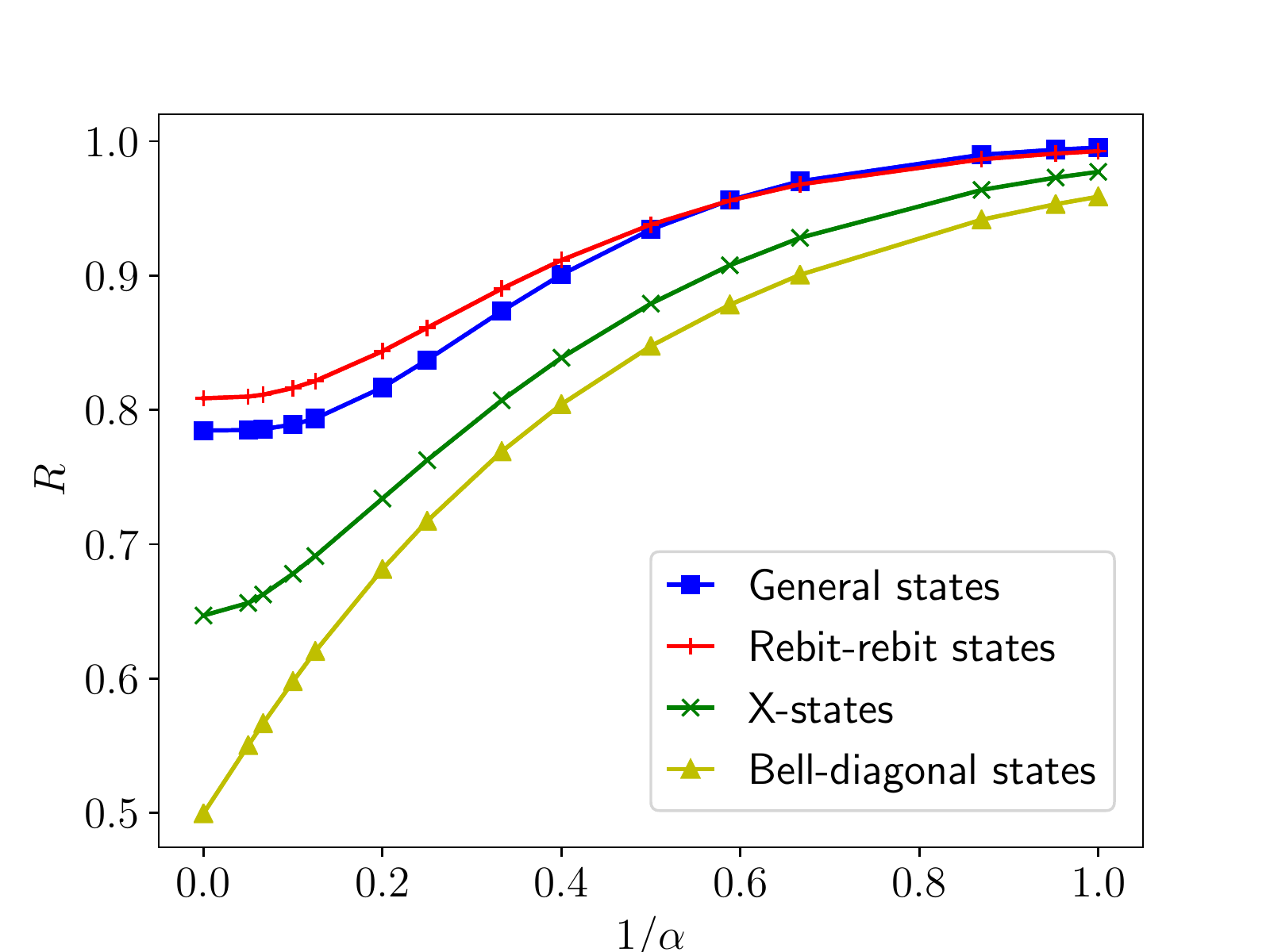}
     \caption{Volume ratios $R$ for the R\'enyi-entropy-based criterion with various $\alpha$. The curves belong to
     different families of qubit--qubit states.}
   \label{fig:22ren}
 \end{figure}

\begin{figure}[h!]
   \centering
  \includegraphics[width=0.5\textwidth]{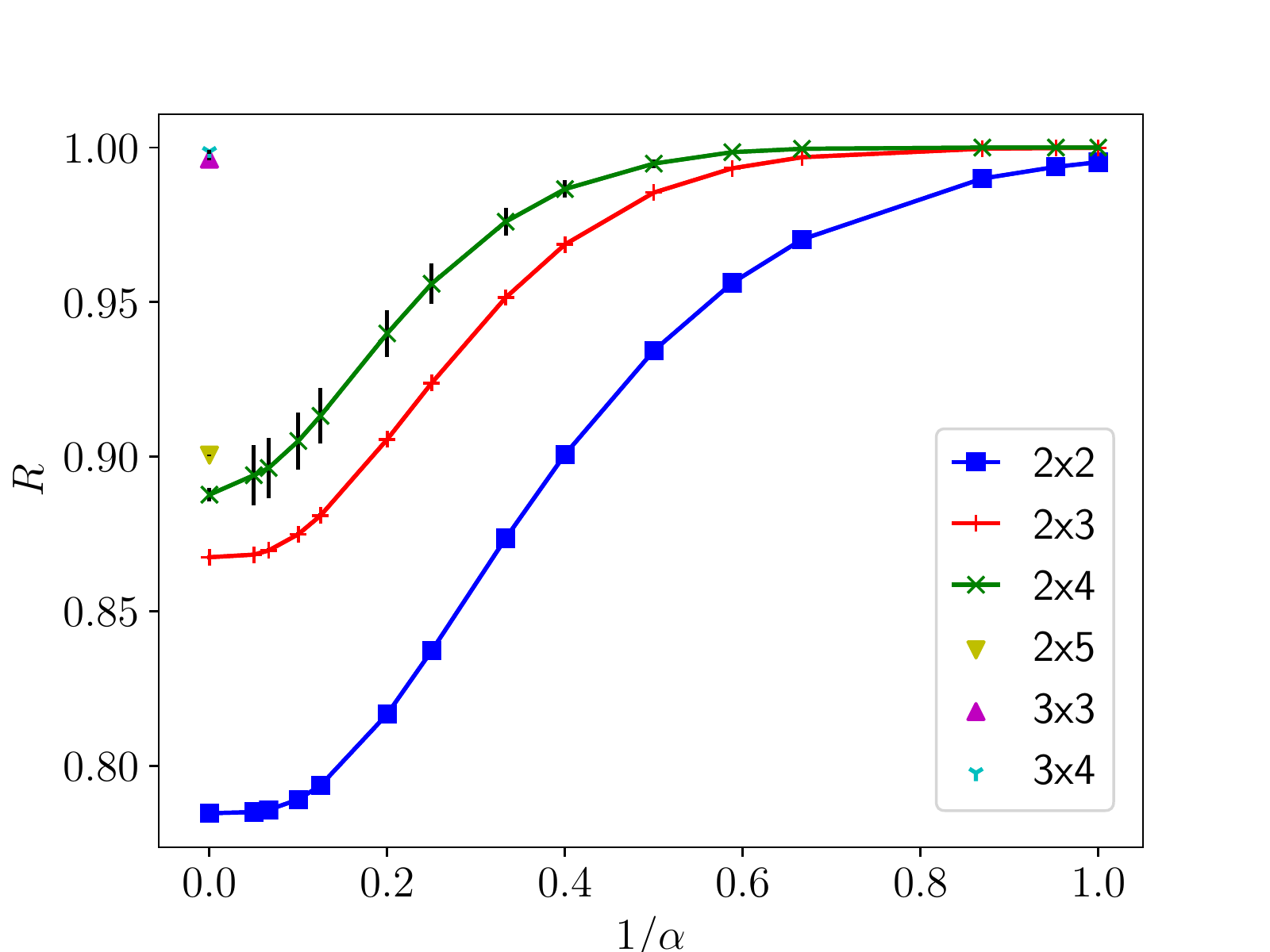}
     \caption{Volume ratios $R$ between quantum states fulfilling R\'enyi-entropy-based criteria and all quantum states. For higher dimensional systems only $\alpha = \infty$ is shown, as this is the most 
     relevant case. For all systems except $2\times 4$ the statistical errors lie within the thickness of the plotted line. }
   \label{fig:allren}
 \end{figure}

The obtained volume ratios for all previously described systems and criteria are listed in Table \ref{tab:ratios}. Additionally, for systems with dimension up to $d=63$, we have investigated the criterion based on 
the R\'enyi entropy for $\alpha \geqslant 1$.
As shown in Figs. \ref{fig:22ren} and \ref{fig:allren}, the criterion based on $S_\infty(\rho)$ yields the lowest estimate for the volume ratio $R$ in all cases.
In fact, this holds not only in average but for every single state. If $\alpha \geqslant \beta$, then a state fulfills the entropy criterion for $S_\alpha(\rho)$, it also fulfills the criterion for $S_\beta(\rho)$.
Therefore, we only list $R$ for $S_\infty(\rho)$ and $S_1(\rho)$ in Table \ref{tab:ratios}, as these volume ratios are the lower and upper bound for $\alpha \geqslant 1$. Here, we need to remind the reader that the 
R\'enyi-entropy-based criteria are defined also for $\alpha \in [0,1)$, but based on our numerical experiences this half-open interval follows the same tendency as what we have observed for $\alpha \geqslant 1$
and therefore in most of the cases our approach is unable to find quantum states which violate this type of criteria. The criterion of $\alpha=1$, i.e., von Neumann entropy, becomes also inconclusive already at 
$2 \times 4$, but we kept it due to its central role in quantum information science.
The sample sizes in the last column of Table \ref{tab:ratios} vary depending on the type of the system. Up to the $3 \times 3$
system $10^7-10^9$ density matrices are enough to obtain accurate estimates, and every sample is generated in a reasonable time, e.g., for the general $2 \times 2$ system with $d=15$ we require a few hours on a laptop to get $10^8$ density matrices. 
However, for larger systems the distances to the borders of the convex body of the density matrices are typically much smaller than the starting interval for any given state and direction, and due to the acceptance-rejection method, the generation of the required sample size takes much longer.
E.g., for the general $3 \times 4$ system with $d=143$ we require more than three weeks on a workstation to obtain $3.5 \times 10^8$ density matrices. 
As we argued in Sec. \ref{III} the sample size has to increase with the dimension of the system to obtain good estimates of the volume ratios, but the time to generate the next density matrix is also increasing with the dimension. 
Therefore, this method with current technologies has its own limitations.

In general, these numerical results may imply an implication arrow from $S_\infty(\rho)$ to $S_\alpha(\rho)$ in Fig. \ref{fig:criteria}, but one has to be careful, because there might be a set of quantum states 
with zero measure, e.g. pure states, which may contradict this observation. However, for now, we leave this numerical observation as a possible conjecture. In Fig. \ref{fig:22ren}, it is interesting to note that the curve
of the rebit--rebit states as a function of $1/\alpha$ has a different behavior than the curves of the other families of quantum states. We believe this is somehow related to the shape of the 
nine-dimensional convex body of rebit--rebit states, where almost half of them are separable; see the analytical ratio $R = 29/64$ obtained by Ref. \cite{Lovas}.

Apart from the previously known dependency, i.e. the equivalence of PPT and the reduction criterion for $2\times N $ systems, there are additional identical values for $R$ apparent in Table \ref{tab:ratios}.
For $2\times 2$ systems, the criteria based on majorization and $S_\infty(\rho)$ are identical, because $\rho_A$ and $\rho_B$ each have two eigenvalues which sum to $1$. 
Thus, only the largest eigenvalue of each matrix is relevant for the majorization criterion.
The same behavior is found for the special $2\times 3$ systems $(i)$ and $(ii)$, although in these cases no such simple explanation exists. In the case of $3\times 3$ and $3\times 4$ systems 
the obtained estimates of $R$ are also identical, however we know that the majorization criterion is more restrictive than $S_\infty(\rho)$. This difference between these criteria has already been observed
for the lower dimensional systems of $2\times 3$ and $2\times 4$. Thus, similar to the least restrictive $S_1(\rho)$ criterion in high dimensions, where the results are inconclusive, 
our algorithm does not find quantum states of $3\times 3$ and $3\times 4$ systems, which fulfill $S_\infty(\rho)$ and violate the majorization criterion within the runtime of our algorithm.
However, the true values of $R$ can still be within one standard deviation of the mean even when the estimates of $R$ are identical.

To shed some light on differences and similarities between these criteria, we are going to investigate mathematically the most easily tractable case, the Bell-diagonal states. Our numerical results in 
Table \ref{tab:ratios} show for Bell-diagonal states that PPT, reduction, majorization, and $S_\infty(\rho)$-based criteria are identical, but the $S_1(\rho)$-based criterion is different. Eq. \eqref{eq:belldiag}
implies that $\rho_A = I_2/2$ and $\rho_B = I_2/2$. Thus, $1/2$ is a degenerate eigenvalue for both subsystems $A$ and $B$. The eigenvalues of $\rho_{\rm BD}$ are
\begin{equation}
  \lambda_{\rm BD,i}=1/4 + f_i(a_x,a_y,a_z),\quad i \in \{1,2,3,4\}, \nonumber
\end{equation}
where
\begin{eqnarray}
 f_1(a_x,a_y,a_z)&=&\frac{1}{2} (-a_x-a_y-a_z), \nonumber \\
 f_2(a_x,a_y,a_z)&=&\frac{1}{2} (a_x+a_y-a_z), \nonumber \\
 f_3(a_x,a_y,a_z)&=&\frac{1}{2} (a_x-a_y+a_z), \nonumber \\
 f_4(a_x,a_y,a_z)&=&\frac{1}{2} (-a_x+a_y+a_z), \nonumber 
\end{eqnarray}
and $a_x,a_y,a_z \in [-0.5,0.5]$ with $\lambda_{\rm BD,i}\geqslant 0$ for all $i$. First, the PPT criterion yields the transformations $a_x \rightarrow a_x$, $a_y \rightarrow -a_y$, and $a_z \rightarrow a_z$ in $\rho_{\rm BD}$, and thus the condition
\begin{equation}
  1/4 - f_i(a_x,a_y,a_z) \geqslant 0, \quad \forall i \in \{1,2,3,4\}. \label{eq:BD-PPT}
\end{equation}
In the case of the reduction criterion, we have
\begin{equation}
 \rho_A \otimes I_{2} -\rho_{\rm BD}= I_{2} \otimes  \rho_B  -\rho_{\rm BD}=\frac{I_4}{2}-\rho_{\rm BD} \nonumber 
\end{equation}
and the condition
\begin{equation}
  1/2-\big[1/4 + f_i(a_x,a_y,a_z)\big] \geqslant 0, \quad \forall i \in \{1,2,3,4\}, \nonumber
\end{equation}
which is equivalent to \eqref{eq:BD-PPT}. As we have mentioned before, for qubit--qubit systems the majorization and $S_\infty(\rho)$-based criteria are identical, where the inequality
\begin{equation}
 \max_i\{\lambda_{\rm BD,i}\} \leqslant \frac{1}{2} \label{eq:BD-Maj}
\end{equation}
has to be fulfilled. It is obvious that \eqref{eq:BD-PPT} and \eqref{eq:BD-Maj} are also equivalent. However, the $S_1(\rho)$-based criterion reads
\begin{equation}
 \ln 2 \leqslant - \sum^4_{i=1} \lambda_{\rm BD,i} \ln \lambda_{\rm BD,i}, \nonumber
\end{equation}
which is very different from \eqref{eq:BD-PPT}. If $a_z=1/3$ and $a_x=-a_y=x$ then $x \in [-5/12,5/12]$. The four equivalent criteria yield that $\rho_{\rm BD}$ is separable when $x \in [-1/12,1/12]$, whereas the 
$S_1(\rho)$-based criterion is fulfilled when $-0.3873 \lesssim x \lesssim 0.3873$, i.e., many entangled states fulfill this criterion. Interestingly, these similarities and differences are also true for the special $2\times 3$ systems $(i)$ and $(ii)$, 
where no simple explanation exists. In general, each criterion results in a different set of inequalities; e.g., the PPT criterion yields inequalities with polynomials, and the R\'enyi entropy-related inequalities 
involve the logarithmic function. Therefore, the mathematical structures of all these criteria are very different, and only in a few special cases are equivalent.

\begin{figure}[h!]
   \centering
  \includegraphics[width=0.5\textwidth]{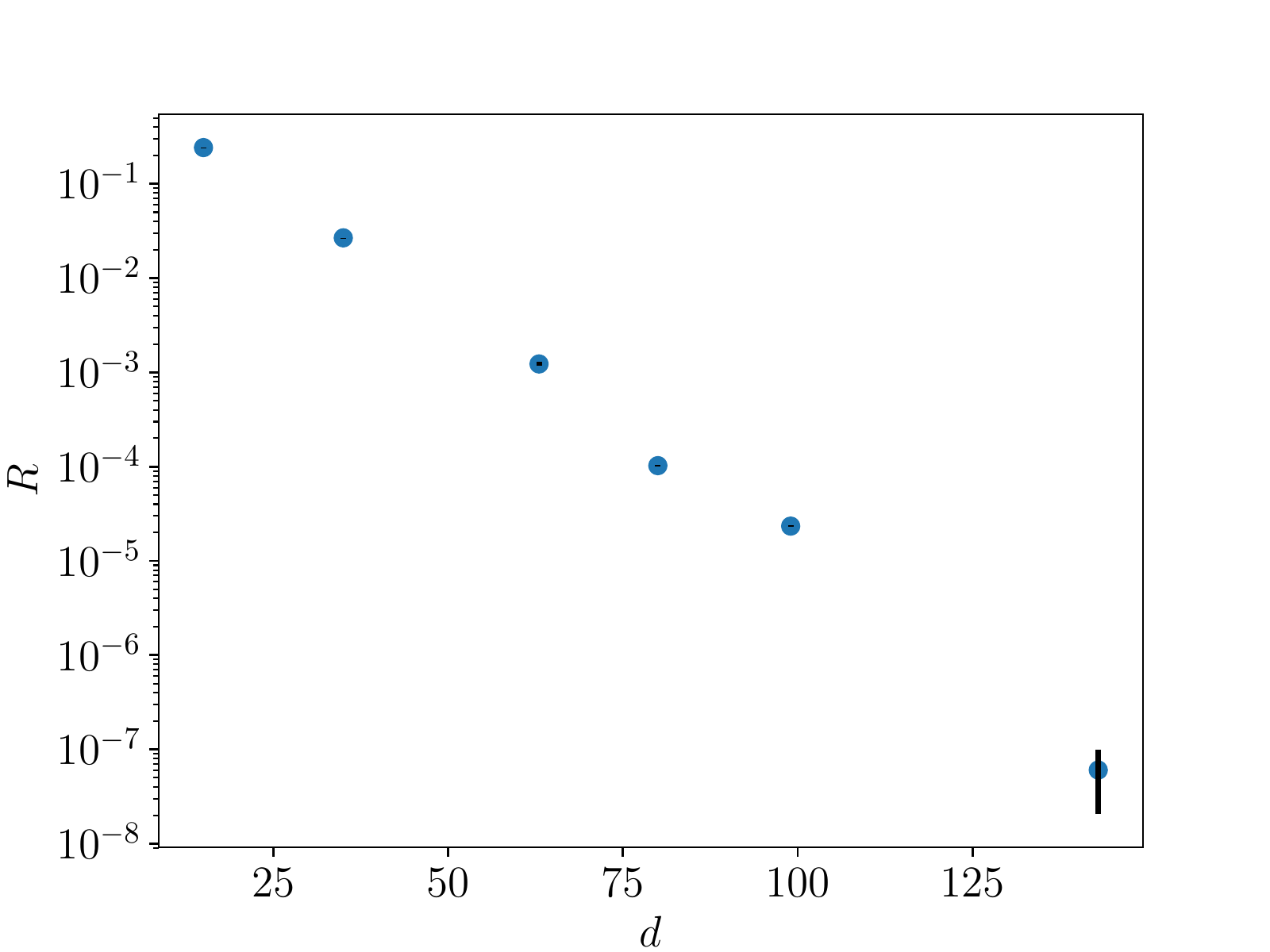}
     \caption{Volume ratios $R$ between PPT and all quantum states for general systems of dimension $d$. For example, $d=35= 2^2 \cdot 3^2 -1$ for qubit--qutrit systems. The statistical errors 
     lie within the thickness of the dots except for $3\times 4$.}
   \label{fig:ppt}
 \end{figure}

Independently from these equivalences or differences between several criteria, we find that the volume ratio between PPT and all quantum states decreases exponentially with the increasing dimension of the 
bipartite quantum system (see Fig. \ref{fig:ppt}). This exponential decrease was first observed numerically in Ref. \cite{ZykLew}, but here we have demonstrated a faster exponential decrease as previously was expected. On the other hand one has to 
take into consideration that PPT states starting from $2 \times 4$ consist of not only separable quantum states but also bound entangled states. Compared to this exponential decrease of volume ratios of PPT states
the other criteria are much less powerful. Furthermore, majorization and entropy-based criteria result in volume ratios which even seemingly converge to $1$. The reduction criterion apart from the $2 \times N$ cases decreases also the 
corresponding volume ratios, but it seems not to have an exponential decrease with increasing dimension of the bipartite quantum system.

\section{Summary and conclusions}
\label{V}

To summarize, we have focused on different entanglement detection criteria and numerically investigated the Euclidean volume ratios between criteria non-violating and all quantum states 
in many bipartite quantum systems. These estimated volume ratios are capable to characterize the performance of every criterion and thus we were able to assign quantitative values to them. 
Our results show that with increasing dimensions only the PPT is the most relevant entanglement detection criterion. The reduction, majorization, and R\'enyi-entropy-based criteria are less 
effective, whereas the last two's performances become worse with the increasing dimension of the bipartite quantum system. For example, in the case of $3 \times 3$ PPT yields at least $99.99\%$ entangled states,
while the majorization criterion suggests that entangled states are around $0.5\%$ of all quantum states. Furthermore, we have also found a hierarchy among the R\'enyi-entropy-based criteria, larger $\alpha$ yields
better entanglement detection. Even though the majorization and R\'enyi-entropy-based criteria are almost useless for large bipartite quantum systems, still maximally entangled states will always violate these 
criteria, which also hints that the volume of these states approaches zero with increasing dimension. Questions concerning the origins of these behaviors may be asked, but usually answers are not that simple if 
one works in high-dimensional Euclidean vector spaces; see, for example, the Busemann-Petty problem \cite{Matousek} for convex bodies symmetric about the origin, like the convex set of all quantum states investigated
in this paper.

Finally, some comments on our numerical method are in order. For $2 \times 5$ and $3 \times 4$ systems the algorithm started to approach its limits in the sense of computational time. The bottleneck of the 
hit-and-run algorithm is to sample enough quantum states such that they have a uniform distribution \cite{Lovasz}. 
With increasing dimension $d$ of the bipartite quantum system around $O(d^4)$ quantum states have to be sampled and for larger $d$ it takes longer to generate and analyze a quantum state, 
which increases the computational time enormously from days to several weeks. Therefore, our method with current technologies
can support future research in low enough dimensional bipartite or multipartite quantum systems.

\begin{acknowledgments}
This work is supported by the Deutsche Forschungsgemeinschaft (DFG) -- SFB 1119 -- 236615297, the DFG under Germany's Excellence Strategy-Cluster of Excellence Matter and Light for Quantum
Computing (ML4Q) EXC 2004/1-390534769, and AIDAS - AI, Data Analytics and Scalable Simulation - which is a Joint Virtual Laboratory gathering
the Forschungszentrum J\"ulich (FZJ) and the French Alternative Energies and Atomic Energy Commission (CEA).
\end{acknowledgments}

\end{document}